# Sign of the Feynman Propagator and Irreversibility

Allan Tameshtit

Ronin Institute, 127 Haddon Pl., Montclair, NJ 07043-2314

For the interacting Feynman propagator $\Delta_{F,\text{int}}(x, y)$ of scalar electrodynamics, we show that the sign property, $\text{Re}\, i\Delta_{F,\text{int}} \geq 0$, may hinge on the reversibility of time evolution. In contrast, $\text{Im}\, i\Delta_{F,\text{int}}$ is indeterminate. When we switch to reduced dynamics under the weak coupling approximation, the positive semidefinite sign of $\text{Re}\, i\Delta_{F,\text{int}}$ is generally lost, unless we impose severe restrictions on the Kraus operators that govern time evolution. With another approximation, the rotating wave approximation, we may recover the sign by restricting the test functions to exponentials under certain conditions.

**Introduction**

The Feynman propagator, $\Delta_F(x, y)$, for charged, scalar particles of mass $m$ is defined by

$$i\Delta_F(x, y) = \langle 0 | T\{\varphi(x)\varphi^\dagger(y)\} | 0 \rangle \tag{1}$$

where $T$ is the time ordering operator, $|0\rangle$ is the vacuum state and $\varphi(x)$ is the field operator,

$$\varphi(x) = \int \frac{1}{2(2\pi)^3 \omega_{\mathbf{k}}} \left( a_{\mathbf{k}} e^{-ikx} + b_{\mathbf{k}}^\dagger e^{ikx} \right) d^3\mathbf{k}, \tag{2}$$

or in manifestly covariant form

$$\varphi(x) = \int \frac{1}{(2\pi)^3} \left( a_{\mathbf{k}} e^{-ikx} + b_{\mathbf{k}}^\dagger e^{ikx} \right) \delta(k^2 - m^2) \theta(k^0) \, d^4\mathbf{k}, \tag{3}$$

where $\theta(k^0)$ is the unit step function and $k^0 = \omega_{\mathbf{k}} = \sqrt{\mathbf{k}^2 + m^2}$ in natural units [1]. In these expressions, the lowering and raising operators for particles and antiparticles satisfy the commutation relations

$$[a_{\mathbf{k}}, a_{\mathbf{k}'}^\dagger] = [b_{\mathbf{k}}, b_{\mathbf{k}'}^\dagger] = (2\pi)^3 2\omega_{\mathbf{k}} \delta^3(\mathbf{k} - \mathbf{k}') \tag{4}$$

with other commutators of these operators vanishing. Interpreted as a Green's function, the Feynman propagator is the solution of the following modified Klein-Gordon equation with a point source [1]

$$\left( \frac{\partial^2}{\partial t^2} - \nabla^2 + m^2 - i\varepsilon \right) \Delta_F(x, y) = -\delta^4(x - y) \tag{5}$$

The provenance of the small, imaginary term (eventually ε→0) is related to the particular choice of boundary conditions that ensures Eq. (1) is the solution of Eq. (5); instead of shifting a contour in the integral solution of Eq. (5) to avoid poles on the x-axis, we instead shift the poles to either side by introducing -iε and elect to maintain the contour on the axis. With the use of Wick's theorem, the Feynman propagator can be used to perturbatively compute many important quantities in quantum field theory, such as scattering cross-sections.

One key property of the Feynman propagator that is used implicitly in such calculations, but not often studied *per se*, relates to its sign, which we will examine in this paper. For this purpose, it is convenient to import some of the bra-ket and operator notation of quantum mechanics, but without some of the interpretations normally associated therewith. This is similar to how Barut uses this notation in a purely classical field theoretic setting ( [2], p. 150-151). Thus, the operator version of Eq. (5) is

$$(-\hat{p}^\mu \hat{p}_\mu + m^2 - i\varepsilon)\widehat{\Delta}_F = -\mathbb{I}. \tag{6}$$

Here, the operator $\hat{p}$ satisfies $\langle x|\hat{p}^\mu|\psi\rangle = \left(i\frac{\partial \psi}{\partial t}, \frac{1}{i}\nabla\psi\right)$ where $\langle x|y\rangle = \delta^4(x-y)$; the kets $|x\rangle$ are not to be interpreted as eigenstates of the position operator. (See [3], section 3c.) Moreover, we should remind ourselves that in quantum field theory, states of the multi-particle system live in Fock space with the ladder and field operators being defined therein. The ket $|x\rangle$, on the other hand, is not a member of Fock space. Likewise, the operator $\hat{p}$ is not to be confused with the field momentum operator **P**.

From Eq. (6), the operator Feynman propagator is $i\widehat{\Delta}_F = i(\hat{p}^\mu \hat{p}_\mu - m^2 + i\varepsilon)^{-1}$. For a more general operator $\hat{\sigma}(\hat{x}, \hat{p})$, the kernel $\sigma(x,y) = \langle x|\hat{\sigma}|y\rangle$ of two space time variables $x$ and $y$ may be used to define a functional

$$\sigma[f] = \langle f|\hat{\sigma}|f\rangle, \tag{7}$$

where

$$\langle f|\hat{\sigma}|g\rangle = \iint f^*(x)\sigma(x,y)g(y)d^4x d^4y. \tag{8}$$

The functional $\sigma$ is positive semidefinite ("positive" for short, when there is no ambiguity) if

$$\iint f^*(x)\sigma(x,y)f(y)d^4x d^4y \geq 0 \tag{9}$$

for all integration regions and test functions $f$ for which the integral exists, and we succinctly write $\sigma \geq 0$. More restrictively, $\sigma$ may be positive for certain regions, such as a specific space-like surface, or a class of test functions of physical relevance. A functional is indeterminate if it has no sign, meaning it is none of positive, positive semidefinite, negative and negative semidefinite. If we let $\hat{\sigma}_1 = \frac{1}{2}(\hat{\sigma} + \hat{\sigma}^\dagger)$ and $\hat{\sigma}_2 = \frac{1}{2i}(\hat{\sigma} - \hat{\sigma}^\dagger)$, then $\hat{\sigma} = \hat{\sigma}_1 + i\hat{\sigma}_2$, and $\langle f|\hat{\sigma}_1|f\rangle$ and $\langle f|\hat{\sigma}_2|f\rangle$ are real numbers,

$$\text{Re} \iint f^*(x)\sigma(x,y)f(y)d^4x d^4y = \iint f^*(x)\sigma_1(x,y)f(y)d^4x d^4y, \tag{10}$$

and

$$\text{Im} \iint f^*(x)\sigma(x,y)f(y)d^4x d^4y = \iint f^*(x)\sigma_2(x,y)f(y)d^4x d^4y. \tag{11}$$

We can write this as $\text{Re}\sigma = \sigma_1$ and $\text{Im}\sigma = \sigma_2$.

It turns out that $\text{Re}\,i\Delta_F \geq 0$. This is relevant, for example, when computing the generating functional for $n$-point functions, which is given by the vacuum-to-vacuum transition amplitude [1], $Z_0[J,J^*]$:

$$Z_0[J,J^*] = \exp\left[-\iint J^*(x)i\Delta_F(x,y)J(y)d^4x\,d^4y\right] \tag{12}$$

where $J(x)$ is a source. That $\text{Re}\,i\Delta_F$ is of positive sign ensures that $Z_0[J,J^*]$ is well defined as a probability amplitude.

Eq. (1) is the Feynman propagator for a free scalar field. The Feynman propagator also arises in the context of interacting fields where it may be defined as

$$i\Delta_{F,\text{int}}(x,y) = \text{Tr}_R\langle 0|T\{\varphi_T(x)\varphi_T^\dagger(y)\}|0\rangle\rho_R \tag{13}$$

where $\rho_R$ is the thermal equilibrium density operator of the electromagnetic field regarded as a heat reservoir (hence the subscript "R") that we trace out. (Cf. Eq. 8.94 of [1].) The field operator $\varphi_T$ in Eq. (13), associated initially with the matter degrees of freedom, are in the Heisenberg picture, evolving under the total (hence the subscript "T") Hamiltonian containing both Klein-Gordon and electromagnetic operators.

For example, use may be made of a thermal Feynman propagator in the study of open quantum electrodynamics [4]. Here, too, the positive sign property of the Feynman propagator is important when calculating relevant amplitudes. Another example where a thermal Feynman propagator is utilized is in the study of the electroweak interaction at high temperatures, in analogy with phase transitions in ferromagnetism [5].

When studying interacting or open systems we may ask whether approximations often employed therein preserve the sign of the Feynman propagator. In this paper, we will show that reversibility---or invertibility of the underlying dynamical map---is central to preserving the aforementioned sign property under such cases. Approximations leading to irreversible dynamics generally destroy the positive sign of $\text{Re}\,i\Delta_{F,\text{int}}$. We will focus on one such approximation, the weak coupling approximation [6], which can yield a completely positive dynamical map,

$$\Lambda_t(\cdot) = \sum_j V_j(t) \cdot V_j^\dagger(t), \tag{14}$$

characterized by Kraus operators $V_j$ satisfying

$$\sum_j V_j^\dagger(t)V_j(t) = \mathbb{1}, \tag{15}$$

the latter equation ensuring that normalization is preserved [7]. The Kraus operators, used for describing the temporal evolution of open systems, are analogous to the unitary evolution operator $U(t)$ that appears in the Liouville propagator $U(t) \cdot U^\dagger(t)$, which is used to describe evolution in isolated systems. For reasons that we will elucidate in detail below, unlike the argument of $U(t)$, which can be positive or negative, it is often necessary to restrict $t$ to non-negative values when dealing with Kraus operators. When $\Lambda_t$ has a group theoretic structure, for instance, we mean the semigroup property ($\Lambda_{t_2}\Lambda_{t_1} = \Lambda_{t_2+t_1}$ for $t_1, t_2 \geq 0$) not the group property. Physically, this means that there is no map $\Lambda_{-t}$ that can reverse the dynamics of $\Lambda_t$. We will show that if we were to insist that $\Lambda(\cdot)$ be invertible then there would exist one Kraus operator chosen from the set $\{V_j\}$ to which all the others would be merely proportional. This restriction is physically untenable for an open system and yet abandoning reversibility jeopardizes the positive sign condition. Fortunately, we will also show that with a further oft-used approximation, the rotating wave approximation [6], we may recover the sign even for reduced dynamics when we restrict ourselves to exponential test functions under certain conditions.

The Feynman propagator, suitably modified, plays a significant role in many quantum field theories. As a concrete example, we will focus on scalar electrodynamics in which the appropriate Schrodinger equation for the scalar field is the Klein-Gordon equation. While admittedly one motivation for concentrating on scalar fields is expediency---they are simpler to treat than fields describing nonzero spin particles---scalar field theories can be used in their own right as effective theories to treat composite particles, such as various mesons subject to nucleon-nucleon interactions [8], and also as a means to treat the fundamental, scalar Higgs boson. In addition, every component of a spin 1/2 and spin 1 field--described by the Dirac equation, and the Maxwell and Proca equations, respectively--also obeys a Klein-Gordon equation [1].

**Sign of the Feynman Propagator**

Using the definition of the Feynman propagator given by Eq. (1) and the identity

$$T\{A^\dagger(t_2)A(t_1)\} = \frac{1}{2}[A^\dagger(t_2), A(t_1)]_+ \\ + \frac{1}{2}[\theta(t_1 - t_2) - \theta(t_2 - t_1)][A(t_1), A^\dagger(t_2)] \quad (16)$$

for any operator $A(t)$, one obtains (cf. Eq. 12.93 of [4])

$$\iint f^*(y) i\Delta_F(x, y) f(x) d^4x d^4y \\ = \frac{1}{2}\iint f^*(x)\langle 0|[\varphi(x), \varphi^\dagger(y)]_+|0\rangle f(y) d^4x d^4y \\ - i\text{Re}\iiint f^*(x)f(y)\theta(x_0 - y_0)\frac{\sin k(x-y)}{(2\pi)^3 \omega_\mathbf{k}} d^3k d^4x d^4y \quad (17)$$

where $k^0 = \omega_\mathbf{k} = \sqrt{\mathbf{k}^2 + m^2}$ and $[A, B]_+ \equiv AB + BA$. Hence, it is immediately apparent that Re$i\Delta_F \geq 0$. The functional Im$i\Delta_F$, associated with the last term on the right-hand side of Eq. (17), will be taken up in Appendix 1. There we will see that Im$i\Delta_F$ is indeterminate.

Eq. (1) may be rewritten as

$$i\Delta_F(x,y) = \text{Tr}[T\{\varphi(x)\varphi^\dagger(y)\}\rho_{\text{vac}}] \tag{18}$$

where $\rho_{\text{vac}} = |0\rangle\langle 0|$. When the charged, scalar field is coupled to the electromagnetic field initially in thermal equilibrium, Eq. (18) suggests a generalization to the following interacting Feynman propagator (cf. Eq. (13))

$$i\Delta_{F,\text{int}}(x,y) = \text{Tr}_T[T\{\varphi_T(x)\varphi_T^\dagger(y)\}\rho_{\text{vac}} \otimes \rho_R] \tag{19}$$

where $\rho_R = \exp(-\beta H_R)/\text{Tr}_R\exp(-\beta H_R)$ is the thermal equilibrium density operator of the electromagnetic field acting as a reservoir. The initial matter field operator $\varphi_T(0, \mathbf{x})$, where the subscript "T" stands for the total system of Klein-Gordon + electromagnetic fields, evolves in the Heisenberg picture according to the total Hamiltonian.

Again using Eq. (16), we may compute

$$\begin{aligned}
i\Delta_{F,\text{int}}(x,y) = {} & \frac{1}{2}\text{Tr}_T\{[\varphi_T(x),\varphi_T^\dagger(y)]_+ \rho_{\text{vac}} \otimes \rho_R\} \\
& + \frac{1}{2}[\theta(x_0 - y_0) - \theta(y_0 - x_0)]\text{Tr}_T\{[\varphi_T(x),\varphi_T^\dagger(y)]\rho_{\text{vac}} \otimes \rho_R\}.
\end{aligned} \tag{20}$$

This equation is similar to Eq. 12.93 of [4], but where in this last reference, electromagnetic fields in the interaction picture appear, instead of matter fields in the Heisenberg picture that appear here.

We can use Eq. (20) to obtain

$$\begin{aligned}
\iint f^*(x) & i\Delta_{F,\text{int}}(x,y)f(y)d^4x d^4y \\
= {} & \frac{1}{2}\iint f^*(x)\text{Tr}_T\left\{[\varphi_T(x),\varphi_T^\dagger(y)]_+ \rho_{\text{vac}} \otimes \rho_R\right\} f(y)d^4x d^4y \\
& + \frac{i}{2}\text{Im}\iint f^*(x)f(y)[\theta(x_0 - y_0) \\
& - \theta(y_0 - x_0)]\text{Tr}_T\{[\varphi_T(x),\varphi_T^\dagger(y)]\rho_{\text{vac}} \otimes \rho_R\}d^4x d^4y
\end{aligned} \tag{21}$$

From the preceding result, we deduce that $\text{Re}\,i\Delta_{F,\text{int}}$ is a positive functional. With reference to the last term of Eq. (21), $\text{Im}\,i\Delta_{F,\text{int}}$ is indeterminate. We show this by looking at the special case where the matter and electromagnetic fields are uncoupled so that $\text{Im}\,i\Delta_{F,\text{int}}$ coincides with $\text{Im}\,i\Delta_F$; in Appendix 1, $\text{Im}\,i\Delta_F$ is shown to be indeterminate.

**Reduced Dynamics under the Weak Coupling Approximation**

An isolated quantum system may be described by a number of dynamical variables that for some purposes is larger than necessary. This number can be reduced by aggregating some of the variables and treating them in only a statistical sense. In this manner, we may derive a smaller open system that is more tractable, albeit subject to residual quantum noise. As an example, scalar electrodynamics relates to spinless charged particles coupled to an electromagnetic field. If the focus is to be the matter field, we may trace out variables pertaining to the electromagnetic field.

For this purpose, we may define a reduced density operator $\rho(t) = \Gamma_{t,t_0}(\varpi)$ evolving according to

$$\Gamma_{t,t_0}(\varpi) = \text{Tr}_R U_T(t,t_0) \varpi \otimes \rho_R U_T^\dagger(t,t_0) \tag{22}$$

where the total propagator $U_T(t,t_0)$ can be obtained from the total Hamiltonian of scalar electrodynamics and, as before, $\rho_R$ is the thermal equilibrium density operator of the electromagnetic field. Assuming at time $t_0$ we somehow prepare the total uncorrelated state $\rho_{t_0} \otimes \rho_R$ consisting of the direct product of the matter and electromagnetic fields, then $\rho_t$ is the reduced density operator at time $t$, where $\rho_{t_0} = \varpi$.

The propagator for the reduced density operator given by Eq. (22) is a specific case of a type of mapping that can be written in the form of Eq. (14), with Eq. (15) ensuring that normalization is preserved. The $V_j(t)$ are known as Kraus operators [7] and those associated with Eq. (22) are given by

$$V_{lm} = {}_R\langle l | U_T(t,t_0) \rho_R^{1/2} | m \rangle_R, \tag{23}$$

the $\{|m\rangle_R\}$ being a complete set of reservoir states.

It is well known that the unitary propagator corresponding to a conservative Hamiltonian possesses three useful properties,

$$\begin{align} U(t_2, t_0) &= U(t_2, t_1) U(t_1, t_0) \quad \text{composition} \tag{24}\\ U(t_1, t_0) &= U(t_1 - t_0, 0) \quad \text{time translation} \tag{25}\\ U(t_2, 0) U(t_1, 0) &= U(t_2 + t_1, 0) \quad \text{group} \tag{26} \end{align}$$

where $t_0, t_1, t_2 \in \mathbb{R}$. Assuming any two of the foregoing properties to be true and $U(t,t) = \mathbb{I}$ implies the third.

Unfortunately, the reduced propagator $\Gamma_{t,t_0}$ appearing in Eq. (22) is not so richly endowed. Although it does inherit time translation invariance so that $\Gamma_{t_1,t_0} = \Gamma_{t_1-t_0,0}$ for $t_1 \geq t_0$ if $H_T$ is conservative, $\Gamma$ does not generally obey the composition rule. Neither does $\Gamma$ generally yield a semigroup ($\Gamma_{t_2,0}\Gamma_{t_1,0} = \Gamma_{t_2+t_1,0}$ for $t_1, t_2 \geq 0$), even if $U_T$ forms a group.

In statistical physics, the composition property is known as the Chapman-Kolmogorov equation, which relies on the Markov approximation [9]. Similarly, for reduced dynamics, one assumption that can be made to recover the composition rule is that the total evolution remains in an uncorrelated state throughout, which is to say [6]

$$U_T(t,t_0) \rho_{t_0} \otimes \rho_R U_T^\dagger(t,t_0) = \Gamma_{t,t_0}(\rho_{t_0}) \otimes \rho_R \tag{27}$$

with $t \geq t_0$. Reduced dynamics given by Eq. (27) would enjoy properties (24)-(26). Because the usefulness of a total Hamiltonian that exactly yields Eq. (27) is doubtful, the assumption of uncorrelated behavior is demoted to an approximation and applied to particular, total systems where appropriate.

In fact, Eq. (27) is in some sense more than is needed in the computation of the interacting Feynman propagator and other multi-time correlation functions, which involve traces over various operators. It is enough, for relevant operators $A_T$ and $B_T$, that [4], [6]

$$\text{Tr}_R[A_T U_T(t,t_0)\rho_{t_0}\otimes\rho_R U_T^\dagger(t,t_0) B_T] \simeq \text{Tr}_R[A_T \Gamma_{t,t_0}(\rho_{t_0})\otimes\rho_R B_T] \quad (28)$$

with $t \geq t_0$, which may be referred to as the weak coupling approximation, an idealization where the reservoir is "refreshed" back to equilibrium when tracing over the reservoir variables in certain coarse-grained expressions. Although the behavior implied by expressions (27) and (28) is a fiction---for an initially uncorrelated total state and coupled degrees of freedom, we would expect to find only one privileged time ($t_0$) at which the system of interest and reservoir are uncorrelated---nevertheless the weak coupling approximation may be reasonable on physical grounds for some operators $A_T$ and $B_T$. (For example, when these last two operators are of the form $A\otimes\mathbb{I}_R$ and $B\otimes\mathbb{I}_R$, with $A$ and $B$ being matter operators, expression (28) is exact.) Attractively, the weak coupling approximation yields

$$\Gamma_{t,t_0}(\varpi) = \text{Tr}_R[U_T(t,t_1)U_T(t_1,t_0)\varpi\otimes\rho_R U_T^\dagger(t_1,t_0)U_T^\dagger(t,t_1)] \quad (29)$$
$$\simeq \text{Tr}_R[U_T(t,t_1)\Gamma_{t_1,t_0}(\varpi)\otimes\rho_R U_T^\dagger(t,t_1)] \quad (30)$$
$$= \Gamma_{t,t_1}[\Gamma_{t_1,t_0}(\varpi)] \quad (31)$$

with $t_0 \leq t_1 \leq t$, where use was made of the composition property (24) of the total propagator in line (29) and approximation (28) with $A_T = B_T^\dagger = U_T(t,t_1)$ in line (30). Hence the weak coupling approximation plays a role in ensuring that the composition rule follows, much like the Markov approximation ensures that the Chapman-Kolmogorov equation is obeyed. Moreover, from the composition property and time translational invariance of the map $\Gamma$, the semigroup property follows.

Multi-time averages were computed in [6], [10] and [11]. We follow [6] in using the weak coupling approximation to obtain multi-time averages below. The adjoint map $\tilde{\Gamma}$, defined by $\text{Tr}[\tilde{\Gamma}(A)\rho] = \text{Tr}[A\Gamma(\rho)]$, together with the weak coupling approximation may be used to make the following replacement:

$$\text{Tr}_T[A\rho_T(t_1)B_T(t_2 - t_1)] \rightarrow \text{Tr}[A\rho(t_1)B(t_2 - t_1)] \quad (32)$$

where $B_T(t) = U_T^\dagger(t)BU_T(t)$, $\rho_T(t) = U_T(t)\rho\otimes\rho_R U_T^\dagger(t)$, $B(t) = \tilde{\Gamma}_t(B)$ and $\rho(t) = \Gamma_t(\rho)$, provided $t_2 \geq t_1$. Operators without subscripts continue to denote system of interest (matter) operators and here $\Gamma_t$ is some dynamical map derived, usually approximately, from Eq. (22). Whereas the time argument in $B_T(t)$ and $\rho_T(t)$ can be positive or negative, the price we pay in moving to reduced dynamics is that the time argument in $B(t)$ and $\rho(t)$ in general must be non-negative, with attendant consequences. We will make repeated use of expression (32) below when computing the interacting Feynman propagator.

**Interacting Feynman Propagator in Reduced Dynamics**

To make use of the substitution (32), it is imperative that we only propagate forward in time. This stipulation at once allows us to obtain reduced dynamics and generally destroys the positive sign of the interacting Feynman propagator.

Later, we will detail why reversibility is not allowed, but for now let us accept this premise and prepare for the application of the weak coupling approximation by expressing a two-time average as follows.

$$\text{Tr}_T[A_T(t_1)\rho\otimes\rho_R B_T(t_2)] \\ = \theta(t_1 - t_2)\text{Tr}_T[A_T(t_1 - t_2)\rho_T(t_2)B] \\ + \theta(t_2 - t_1)\text{Tr}_T[A\rho_T(t_1)B_T(t_2 - t_1)] \quad (33)$$

where we have assumed the underlying total Hamiltonian is conservative and $t_1, t_2 \geq 0$. (If Eq. (33) is to apply when $t_1 = t_2$, we must define $\theta(0) = 1/2$.) Only one of the two terms on the right-hand side is non-zero. Because in that surviving term only non-negative time arguments appear, we may use substitution (32) in the last term and a similar substitution in the first term on the right-hand side of Eq. (33) to yield

$$\text{Tr}_T[A_T(t_1)\rho\otimes\rho_R B_T(t_2)] \to \theta(t_1 - t_2)\text{Tr}[A(t_1 - t_2)\rho(t_2)B] \\ + \theta(t_2 - t_1)\text{Tr}[A\rho(t_1)B(t_2 - t_1)]. \quad (34)$$

Even though $A(t_1 - t_2)$ is not defined if $t_1 < t_2$, we assume that in such case the step function dominates and take $\theta(t_1 - t_2)\text{Tr}[A(t_1 - t_2)\rho(t_2)B]$ to be zero, and similarly for the last term if $t_2 < t_1$. Using the substitution (34) in Eq. (21), we may approximate

$$\iint f^*(x)i\Delta_{F,\text{int}}(x,y)f(y)d^4x d^4y \cong \\ \text{Re}\iint f^*(x)f(y)\theta(x_0 - y_0)\text{Tr}\{[\varphi^\dagger(0,\mathbf{y}),\phi(x_0 - y_0,\mathbf{x})]_+\rho_{\text{vac}}(y_0)\}d^4x d^4y \quad (35) \\ +i\text{Im}\iint f^*(x)f(y)\theta(x_0 - y_0)\text{Tr}\{[\phi(x_0 - y_0,\mathbf{x}),\varphi^\dagger(0,\mathbf{y})]\rho_{\text{vac}}(y_0)\}d^4x d^4y$$

Although the exact $\text{Re}i\Delta_{F,\text{int}}$ is a positive functional, after invoking the weak coupling approximation we see from the first term of the right-hand side of expression (35) that we have manifestly lost the sign property. If the dynamics were reversible, which we will see in the next section it generally is not, we could reconstitute the sign of $\text{Re}i\Delta_{F,\text{int}}$ from expression (35).

To wit, suppose $\tilde{\Gamma}_t$ possessed the group property, were invertible and satisfied (cf. equation 2.3 of [12])

$$[\tilde{\Gamma}_t(A)]^\dagger = \tilde{\Gamma}_t(A^\dagger) \quad (36)$$

and

$$\tilde{\Gamma}_t(AB) = \tilde{\Gamma}_t(A)\tilde{\Gamma}_t(B). \quad (37)$$

(In fact, Eqs. (36) and (37) follow from invertibility, as characterized in Appendix 2, and are therefore superfluous.) With these properties in hand, we would compute

$$\text{Tr}\{[\varphi^\dagger(\mathbf{y}), \phi(x_0 - y_0, \mathbf{x})]_+ \rho_{\text{vac}}(y_0)\} = \text{Tr}\left\{\tilde{\Gamma}_{y_0}[\varphi^\dagger(\mathbf{y}), \tilde{\Gamma}_{-y_0}\tilde{\Gamma}_{x_0}\phi(\mathbf{x})]_+ \rho_{\text{vac}}\right\} \quad (38)$$
$$= \langle 0|[\varphi^\dagger(y), \varphi(x)]_+|0\rangle. \quad (39)$$

That we used invertibility is apparent from the presence of $\tilde{\Gamma}_{-y_0}$ in Eq. (38). Substituting Eq. (39) into the first term on the right-hand side of expression (35), we would obtain

$$\text{Re}\iint f^*(x) i\Delta_{F,\text{int}}(x,y) f(y) d^4x d^4y$$
$$\cong \text{Re}\iint f^*(x) f(y) \theta(x_0 - y_0)\langle 0|[\varphi^\dagger(y), \varphi(x)]_+|0\rangle d^4x d^4y \quad (40)$$

$$= \tfrac{1}{2}\iint f^*(x) f(y)\langle 0|[\varphi^\dagger(y), \varphi(x)]_+|0\rangle d^4x d^4y. \quad (41)$$

As a reminder, we note that unlike in Eq. (17), the field operators in Eq. (41), for example, are evolving under reduced dynamics, $\varphi(x) = \tilde{\Gamma}_{x_0}[\varphi(\mathbf{x})]$. Again we see that $\text{Re}i\Delta_{F,\text{int}} \geq 0$. Thus, pretending that reduced dynamics were invertible allows us to recover the positive sign.

**Irreversibility of Reduced Dynamics**

We have seen that the positive sign of $\text{Re}i\Delta_F$ may hinge on the invertibility of the dynamical map, so it is important to understand where the notion of irreversibility of reduced dynamics comes from.

If $\mathcal{B}_1(\mathcal{H})$ denotes the Banach space of trace class operators $\sigma$ on the separable Hilbert space $\mathcal{H}$ with trace norm $\|\sigma\|_1 = \text{Tr}\sqrt{\sigma^\dagger \sigma}$, then $\bar{\mathcal{B}}^+(\mathcal{H}) \subset \mathcal{B}_1(\mathcal{H})$ is the set of physical density operators, which are positive, self-adjoint and normalized ($\|\rho\|_1 = 1, \rho \in \bar{\mathcal{B}}^+(\mathcal{H})$) [12], [13], [14]. On physical grounds, it has been argued that physical density operators should be completely positive, not merely positive [15]. For our purposes, completely positive operators can be taken to mean operators that arise from evolution governed by Eq. (14). (A frequently cited example of an operator that is positive, but not completely positive, is the operator that returns the transpose of the density matrix [16].)

A completely positive dynamical map $\Lambda_t$ depends on a time parameter and maps $\rho \in \bar{\mathcal{B}}^+(\mathcal{H})$ to $\sum_j V_j(t)\rho V_j^\dagger(t) \in \bar{\mathcal{B}}^+(\mathcal{H})$ where the $V_j$ are operators on $\mathcal{H}$ that satisfy the normalization condition, Eq. (15). Such maps are convex linear [4], [12] meaning that

$$\Lambda(p\rho_1 + (1-p)\rho_2) = p\Lambda(\rho_1) + (1-p)\Lambda(\rho_2) \quad (42)$$

where $\rho_1, \rho_2 \in \bar{\mathcal{B}}^+$ and $0 \leq p \leq 1$.

To understand why we must insist on a preferred direction of time, consider the following example of a manifestly covariant, completely positive dynamical map (cf. equation 5.20 of [17]),

$$e^{-\tau\{\lambda^\mu \hat{p}_\mu,\cdot,\lambda^\nu \hat{p}_\nu\}}\rho = \frac{1}{2\sqrt{\pi\tau}} \int_{-\infty}^{\infty} \exp\left(-\frac{u^2}{4\tau}\right) e^{-i\lambda^\mu \hat{p}_\mu u} \rho e^{i\lambda^\nu \hat{p}_\nu u} du \qquad (43)$$

where $\tau$ is a scalar playing the role of proper time, $\lambda^\mu$ are real 4-vector parameters, $\hat{p}^\mu$ is the four-momentum Hermitian operator and

$$\{A, \rho, B\} \equiv BA^\dagger \rho + \rho BA^\dagger - 2A^\dagger \rho B. \qquad (44)$$

This dynamical map gives rise to the master equation

$$\frac{d\rho}{d\tau} = -\{\lambda^\mu \hat{p}_\mu, \rho(\tau), \lambda^\nu \hat{p}_\nu\}, \qquad (45)$$

which is in prototypical Lindblad form [15]. A candidate for an inverse is [18]

$$e^{\tau\{\lambda^\mu \hat{p}_\mu,\cdot,\lambda^\nu \hat{p}_\nu\}}\rho = \frac{1}{2\sqrt{\pi\tau}} \int_{-\infty}^{\infty} \exp\left(-\frac{u^2}{4\tau}\right) e^{\lambda^\mu \hat{p}_\mu u} \rho e^{-\lambda^\nu \hat{p}_\nu u} du, \qquad (46)$$

which, by direct computation, can be seen to satisfy

$$e^{\tau\{\lambda^\mu \hat{p}_\mu,\cdot,\lambda^\nu \hat{p}_\nu\}} e^{-\tau\{\lambda^\mu \hat{p}_\mu,\cdot,\lambda^\nu \hat{p}_\nu\}}\rho = \rho. \qquad (47)$$

What then is the problem with claiming that $e^{\tau\{\lambda^\mu \hat{p}_\mu,\cdot,\lambda^\nu \hat{p}_\nu\}}$ is the inverse of $e^{-\tau\{\lambda^\mu \hat{p}_\mu,\cdot,\lambda^\nu \hat{p}_\nu\}}$ and therefore that the latter exhibits reversible dynamics? A priori, there are three possible reasons that could preclude invertibility: $e^{-\tau\{\lambda^\mu \hat{p}_\mu,\cdot,\lambda^\nu \hat{p}_\nu\}}$ is a) not one-to-one on, but is onto $\overline{\mathcal{B}}^+$, b) one-to-one on, but not onto $\overline{\mathcal{B}}^+$, or c) neither one-to-one on nor onto $\overline{\mathcal{B}}^+$.

Suppose

$$0 = e^{-\tau\{\lambda^\mu \hat{p}_\mu,\cdot,\lambda^\nu \hat{p}_\nu\}}\rho_1 - e^{-\tau\{\lambda^\mu \hat{p}_\mu,\cdot,\lambda^\nu \hat{p}_\nu\}}\rho_2 \qquad (48)$$

$$= \iint \langle p|(\rho_1 - \rho_2)|p'\rangle e^{-[\lambda^\mu(p_\mu - p'_\mu)]^2 \tau} |p\rangle\langle p'| d^4p\, d^4p' \qquad (49)$$

If the $|p\rangle\langle p'|$ are linearly independent, we require $\langle p|\rho_1|p'\rangle = \langle p|\rho_2|p'\rangle$, which implies $\rho_1 = \rho_2$. I.e., $e^{-\tau\{\lambda^\mu \hat{p}_\mu,\cdot,\lambda^\nu \hat{p}_\nu\}}$ is one-to-one. Hence, the problem must be that $\Lambda_\tau \equiv e^{-\tau\{\lambda^\mu \hat{p}_\mu,\cdot,\lambda^\nu \hat{p}_\nu\}}$ is not onto $\overline{\mathcal{B}}^+$. Let us confirm this.

Because $e^{-\tau\{\lambda^\mu \hat{p}_\mu,\cdot,\lambda^\nu \hat{p}_\nu\}}: \overline{\mathcal{B}}^+ \to \text{Ran}\Lambda_\tau \subset \overline{\mathcal{B}}^+$ is one-to-one, the inverse $e^{\tau\{\lambda^\mu \hat{p}_\mu,\cdot,\lambda^\nu \hat{p}_\nu\}}: \text{Ran}\Lambda_\tau \to \overline{\mathcal{B}}^+$ exists and is given by Eq. (46). We wish to show that $\text{Ran}\Lambda_\tau$ is a proper subset of $\overline{\mathcal{B}}^+$. For every pure state $|\psi\rangle\langle\psi| \in \overline{\mathcal{B}}^+$, we have

$$e^{-\tau\{\lambda^\mu \hat{p}_\mu,\cdot,\lambda^\nu \hat{p}_\nu\}}|\psi\rangle\langle\psi| = \frac{1}{2}\sigma_{\psi,-}(\tau) + \frac{1}{2}\sigma_{\psi,+}(\tau) \qquad (50)$$

where $\sigma_{\psi,\pm}(\tau) \in \overline{\mathcal{B}}^+$ with

$$\sigma_{\psi,\pm}(\tau) = \frac{1}{\sqrt{\pi\tau}} \int_0^\infty \exp\left(-\frac{u^2}{4\tau}\right) e^{\mp i\lambda^\mu \hat{p}_\mu u} |\psi\rangle\langle\psi| e^{\pm i\lambda^\nu \hat{p}_\nu u} du. \qquad (51)$$

In the position representation,

$$\langle x|\sigma_{\psi,\pm}(\tau)|x'\rangle = \frac{1}{\sqrt{\pi\tau}} \int_0^\infty \exp\left(-\frac{u^2}{4\tau}\right) \psi(x^\mu \pm \lambda^\mu u)\, \psi^*(x'^\nu \pm \lambda^\nu u)\, du. \qquad (52)$$

The density operators $\sigma_{\psi,+}(\tau)$ and $\sigma_{\psi,-}(\tau)$ are statistical mixtures consisting of a Gaussian-weighted sum of pure states $|\psi\rangle\langle\psi|$ uniformly translated $\lambda^\mu u$ to the left and right, respectively, the Gaussian having a standard deviation of $\sqrt{2\tau}$.

Expressing $|\psi\rangle$ as a superposition of plane waves, $|\psi\rangle = \int_{-\infty}^{\infty} \tilde{\psi}(p)|p\rangle d^4p$, and taking an arbitrary physical state $|\phi\rangle$, we compute

$$\langle\phi|e^{\tau\{\lambda^\mu \hat{p}_\mu, \lambda^\nu \hat{p}_\nu\}}\sigma_{\psi,\pm}|\phi\rangle$$
$$= |\langle\phi|\psi\rangle|^2$$
$$\pm \iint_{-\infty}^{\infty} \text{Im}\{\text{erf}[i\lambda^\mu(p_\mu - p'_\mu)\sqrt{\tau}]\}\text{Im}[\tilde{\psi}(p)\tilde{\phi}^*(p)\tilde{\psi}^*(p')\tilde{\phi}(p')]d^4p d^4p' \quad (53)$$

If we choose $\phi$ to be orthogonal to $\psi$ and assume the real, last term on the right-hand side of Eq. (53) to be non-vanishing, then either $\langle\phi|e^{\tau\{\lambda^\mu \hat{p}_\mu, \lambda^\nu \hat{p}_\nu\}}\sigma_{\psi,-}|\phi\rangle$ or $\langle\phi|e^{\tau\{\lambda^\mu \hat{p}_\mu, \lambda^\nu \hat{p}_\nu\}}\sigma_{\psi,+}|\phi\rangle$ is negative, which implies that either $\sigma_{\psi,-}$ or $\sigma_{\psi,+}$ is not in the domain of $e^{\tau\{\lambda^\mu \hat{p}_\mu, \lambda^\nu \hat{p}_\nu\}}$ despite both being physical density operators. This means that there is no pair $\rho_\pm \in \bar{\mathcal{B}}^+$ such that $e^{-\tau\{\lambda^\mu \hat{p}_\mu, \lambda^\nu \hat{p}_\nu\}}\rho_\pm = \sigma_{\psi,\pm}$, which is to say $e^{-\tau\{\lambda^\mu \hat{p}_\mu, \lambda^\nu \hat{p}_\nu\}}$ is not onto $\bar{\mathcal{B}}^+$. The map $e^{\tau\{\lambda^\mu \hat{p}_\mu, \lambda^\nu \hat{p}_\nu\}}$ is not defined on the set of all physical states. A result along these lines was reported in [18].

There is another way to arrive at this conclusion using an argument presented in Theorem 3.4.1 of [13]. Suppose a completely positive dynamical map $\Lambda: \bar{\mathcal{B}}^+ \to \bar{\mathcal{B}}^+$ takes a pure state to a mixture, $|\psi\rangle\langle\psi| \mapsto p\rho_1 + (1-p)\rho_2$, where $0 < p < 1$ and the physical states $\rho_1$ and $\rho_2$ are unequal. Then a convex linear inverse $\Lambda^{-1}: \text{Ran}\Lambda \to \bar{\mathcal{B}}^+$ cannot be defined at both $\rho_1$ and $\rho_2$ for otherwise $\rho_i = \Lambda\Lambda^{-1}(\rho_i) = \Lambda(|\psi\rangle\langle\psi|) = p\rho_1 + (1-p)\rho_2$, $i=1,2$, which is a contradiction. (The second equation follows because $p\Lambda^{-1}(\rho_1) + (1-p)\Lambda^{-1}(\rho_2) = |\psi\rangle\langle\psi| \Rightarrow \Lambda^{-1}(\rho_i) = |\psi\rangle\langle\psi|$.)

In our above example, Eq. (45) yields, for an initial pure state,

$$\frac{d\text{Tr}\rho^2}{d\tau}\Big|_{\tau=0} = -4\Delta^2(\lambda^\mu \hat{p}_\mu) \quad (54)$$

where the initial uncertainty $\Delta(\lambda^\mu \hat{p}_\mu) = \sqrt{\langle(\lambda^\mu \hat{p}_\mu)^2\rangle - \langle\lambda^\mu \hat{p}_\mu\rangle^2}$. Since $\text{Tr}\rho^2 < 1$ only if $\rho$ is a statistical mixture, for any pure state with $\Delta(\lambda^\mu \hat{p}_\mu) \neq 0$, a pure state maps to a mixture for short enough values of $\tau$. Consequently, the domain of $e^{\tau\{\lambda^\mu \hat{p}_\mu, \lambda^\nu \hat{p}_\nu\}}$ cannot be all of $\bar{\mathcal{B}}^+$ for such values of $\tau$. In some sense, this is not surprising since the map given by Eq. (46) is not manifestly positive; recalling that either $\sigma_{\psi,-}$ or $\sigma_{\psi,+}$ is not in the domain of $e^{\tau\{\lambda^\mu \hat{p}_\mu, \lambda^\nu \hat{p}_\nu\}}$, this map yields a positive image for some, but not all physical states.

For a more general dynamical map $\Lambda$, it is shown in Appendix 2 that invertibility imposes severe restrictions on the associated Kraus operators. Under reasonable hypotheses, if the map given by Eq. (14) is invertible, then there exists one Kraus operator chosen from the set $\{V_j\}$ to which all the others are merely proportional (cf. Theorem 3.4.1 in [13] and Theorem 6.39 in [19]). Except in this extreme case, we cannot assume that $\Lambda$ is invertible, and therefore cannot perform the steps at the end of the last section to demonstrate that $\text{Re}i\Delta_{F,\text{int}} \geq 0$.

**Rotating Wave Approximation and Recovery of Positive Sign**

If we make a further approximation, the rotating wave approximation, we may recover a more modest type of positive sign condition, viz., $\text{Re}i\Delta_{F,\text{int}} \geq 0$ for exponential test functions,

$$f(x) = e^{-\bar{\omega}t}h(\mathbf{x}) \tag{55}$$

with $\bar{\omega} > 0$ and $t \geq 0$. This result is not just academic. Such exponential functions arise, for example, when using adiabatic switching methods to derive the Gell-mann-Low theorem and reduction formulae that follow therefrom [20]. Sufficient conditions for a positive sign with these exponential functions are given below by inequality (67).

In the field of quantum optics, the rotating wave approximation is used to obtain the well-known quantum optical master equation [21]

$$\frac{d\rho}{dt} = \frac{1}{\hbar i}[\hbar \omega a^\dagger a, \rho] - \frac{\gamma(\omega)}{2}[n(\omega)\{a, \rho, a\} + (n(\omega) + 1)\{a^\dagger, \rho, a^\dagger\}]. \tag{56}$$

The rotating wave approximation effectively replaces coupling terms $(a + a^\dagger)(a_j + a_j{}^\dagger)$ between the system and reservoir position variables by $aa_j{}^\dagger + a^\dagger a_j$. The justification for this approximation is typically provided in the interaction picture where rapidly oscillating terms are ignored when averaging over relevant time scales. For more details, Refs. [4], [6] and [21] may be consulted. By analogy to Eq. (56), we can consider the following master equation for scalar electrodynamics

$$\begin{aligned}\frac{d\rho}{dt} &= \frac{1}{2(2\pi)^3 i}\int [a_\mathbf{k}^\dagger a_\mathbf{k} + b_\mathbf{k}^\dagger b_\mathbf{k}, \rho]d^3k \\ &- \int \{h_{11}(\mathbf{k})a_\mathbf{k}, \rho, h_{11}(\mathbf{k})a_\mathbf{k}\}d^3k \\ &- \int \{h_{12}(\mathbf{k})a_\mathbf{k}^\dagger, \rho, h_{12}(\mathbf{k})a_\mathbf{k}^\dagger\}d^3k \\ &- \int \{h_{21}b_\mathbf{k}, \rho, h_{21}(\mathbf{k})b_\mathbf{k}\}d^3k - \int \{h_{22}(\mathbf{k})b_\mathbf{k}^\dagger, \rho, h_{22}(\mathbf{k})b_\mathbf{k}^\dagger\}d^3k\end{aligned} \tag{57}$$

where the raising and lowering operators satisfy Eq. (4). A more realistic master equation would couple particles and anti-particles with terms of the form $\{a + b^\dagger, \cdot, a + b^\dagger\}$ and $\{a^\dagger + b, \cdot, a^\dagger + b\}$. In this vein, see also [22]. Because our primary goal in this section is to demonstrate how physical approximations can lead to a recovery of the positive sign that was lost due to irreversibility, we will content ourselves with Eq. (57).

In adjoint form,

$$\frac{dO}{dt} = \tilde{L}(O)$$

$$= -i[O, H] + \sum_{i,j=1}^{2} \tilde{L}_{ij}(O) \tag{58}$$

where

$$H = \frac{1}{2(2\pi)^3} \int (a_{\mathbf{k}}^\dagger a_{\mathbf{k}} + b_{\mathbf{k}}^\dagger b_{\mathbf{k}}) d^3\mathbf{k},$$

$$\tilde{L}_{11}(\cdot) = \int |h_{11}(\mathbf{k})|^2 a_{\mathbf{k}}[\cdot, a_{\mathbf{k}}^\dagger] d^3k + \int |h_{11}(\mathbf{k})|^2 [a_{\mathbf{k}}, \cdot] a_{\mathbf{k}}^\dagger d^3k,$$

$$\tilde{L}_{12}(\cdot) = \int |h_{12}(\mathbf{k})|^2 a_{\mathbf{k}}^\dagger[\cdot, a_{\mathbf{k}}] d^3k + \int |h_{12}(\mathbf{k})|^2 [a_{\mathbf{k}}^\dagger, \cdot] a_{\mathbf{k}} d^3k,$$

$$\tilde{L}_{21}(\cdot) = \int |h_{21}(\mathbf{k})|^2 b_{\mathbf{k}}[\cdot, b_{\mathbf{k}}^\dagger] d^3k + \int |h_{21}(\mathbf{k})|^2 [b_{\mathbf{k}}, \cdot] b_{\mathbf{k}}^\dagger d^3k,$$

and

$$\tilde{L}_{22}(\cdot) = \int |h_{22}(\mathbf{k})|^2 b_{\mathbf{k}}^\dagger[\cdot, b_{\mathbf{k}}] d^3k + \int |h_{22}(\mathbf{k})|^2 [b_{\mathbf{k}}^\dagger, \cdot] b_{\mathbf{k}} d^3k.$$

By direct computation, we can verify the following commutators

$$[[\cdot, H], \tilde{L}_{i1} + \tilde{L}_{i2}] = [\tilde{L}_{11} + \tilde{L}_{12}, \tilde{L}_{21} + \tilde{L}_{22}] = 0, \quad i = 1, 2$$

and hence the solution of the adjoint Eq. (58) is

$$\tilde{\Gamma}_t(O) = e^{t\tilde{L}}(O)$$
$$= e^{-ti[\cdot, H]} e^{t(\tilde{L}_{11} + \tilde{L}_{12})} e^{t(\tilde{L}_{21} + \tilde{L}_{22})}(O).$$

Factoring the propagator in this way makes it easier to compute the following multi-time averages

$$\left\langle 0 \left| \tilde{\Gamma}_\tau \left( C_{ij}(\mathbf{k}', t') C_{lm}(\mathbf{k}'', t'') \right) \right| 0 \right\rangle$$
$$= \delta_{il}(1 - \delta_{mj})\delta^3(\mathbf{k}' - \mathbf{k}'')(2\pi)^3 2\omega_{\mathbf{k}'} e^{\chi_l(\mathbf{k}',\mathbf{k}',t',t'')} e^{i\omega_{\mathbf{k}'}[t'(1-2\delta_{1j})+t''(1-2\delta_{1m})]}$$
$$\times \begin{pmatrix} (1 - \delta_{2j})(1 - \delta_{1m}) e^{i\omega_{\mathbf{k}'}[(1-2\delta_{1j})+(1-2\delta_{1m})]\tau} e^{\chi_l(\mathbf{k}',\mathbf{k}',\tau,\tau)} \\ +4(2\pi)^3 |h_{lm}(\mathbf{k}')|^2 \omega_{\mathbf{k}'} \tau \dfrac{e^{\chi_l(\mathbf{k}',\mathbf{k}',\tau,\tau)} - 1}{\chi_l(\mathbf{k}', \mathbf{k}', \tau, \tau)} \end{pmatrix} \tag{59}$$

where we have defined $C_{11} = C_{12}^\dagger = a$, $C_{21} = C_{22}^\dagger = b$ and

$$\chi_l(\mathbf{k}', \mathbf{k}'', t', t'') =$$
$$(2\pi)^3 2[(|h_{l1}(\mathbf{k}')|^2 - |h_{l2}(\mathbf{k}')|^2)\omega_{\mathbf{k}'} t' + (|h_{l1}(\mathbf{k}'')|^2 - |h_{l2}(\mathbf{k}'')|^2)\omega_{\mathbf{k}''} t''].$$

The $C_{ij}(\mathbf{k},t) = \tilde{\Gamma}_t[C_{ij}(\mathbf{k})]$ are evolving under reduced dynamics. For other models, multi-time averages of the form

$$\langle 0|\tilde{\Gamma}_\tau[C_{ij}(\mathbf{k}',t')C_{lm}(\mathbf{k}'',t'')]|0\rangle = \delta_{il}(1-\delta_{jm})\delta^3(\mathbf{k}'-\mathbf{k}'')g_{ijlm}(t',t'',\tau,\mathbf{k}',\mathbf{k}''), \quad (60)$$

whatever the specific values of the c-functions $g_{ijmn}$, may only be approximately valid. For our purposes, expression (60) will be referred to as the rotating wave approximation keeping in mind that for the master equation (57), the multi-time averages are given exactly by the right-hand side of expression (60), with the $g_{ijlm}$ gleaned from Eq. (59).

Let us consider the test functions given by Eq. (55) and use approximation (35), which together with the convolution theorem yields:

$$\text{Re}\iint f^*(y) i\Delta_{F,\text{int}}(x,y) f(x) d^4x d^4y$$
$$\cong \text{Re}\iint h^*(\mathbf{x})h(\mathbf{y})\theta(t_1)\theta(t_2) e^{-\bar{\omega}(t_1+2t_2)} \text{Tr}\left([\varphi(t_1,\mathbf{x}),\varphi^\dagger(0,\mathbf{y})]_+ \rho_{\text{vac}}(t_2)\right) d^4x d^4y. \quad (61)$$

The field operator is given by

$$\varphi(x) = \int \frac{1}{2(2\pi)^3 \omega_\mathbf{k}} \left(a_\mathbf{k}(t)e^{i\mathbf{k}\cdot\mathbf{x}} + b_\mathbf{k}^\dagger(t)e^{-i\mathbf{k}\cdot\mathbf{x}}\right) d^3k \quad (62)$$

where $\omega_\mathbf{k} = \sqrt{\mathbf{k}^2 + m^2}$, and $a_\mathbf{k}(t) = \tilde{\Gamma}_t(a_\mathbf{k})$ and $b_\mathbf{k}(t) = \tilde{\Gamma}_t(b_\mathbf{k})$ are the particle and antiparticle lowering operators evolving according to reduced dynamics.

We may then compute

$$\text{Re}\iint f^*(x) i\Delta_{F,\text{int}}(x,y) f(y) d^4x d^4y$$
$$\to \text{Re}\int d^3x \int d^3y\, h^*(\mathbf{x})h(\mathbf{y}) \int_0^\infty dt_1 \int_0^\infty dt_2\, e^{-\bar{\omega}(t_1+2t_2)} \int \frac{d^3k}{(2\pi)^3 2\omega_\mathbf{k}} \int \frac{d^3k'}{(2\pi)^3 2\omega_{\mathbf{k}'}} \quad (63)$$
$$\times \langle 0|\tilde{\Gamma}_{t_2}\left([e^{i\mathbf{k}\cdot\mathbf{x}}a_\mathbf{k}(t_1) + e^{-i\mathbf{k}\cdot\mathbf{x}}b_\mathbf{k}^\dagger(t_1), e^{-i\mathbf{k}'\cdot\mathbf{y}}a_{\mathbf{k}'}^\dagger(0) + e^{i\mathbf{k}'\cdot\mathbf{y}}b_{\mathbf{k}'}(0)]_+\right)|0\rangle$$

from which we see that $\text{Re}\, i\Delta_{F,\text{int}}$ is not generally a positive functional under the weak coupling approximation.

However, if we now employ the rotating wave approximation, we get

$$\text{Re}\iint f^*(x) i\Delta_{F,\text{int}}(x,y) f(y) d^4x d^4y$$
$$= \text{Re}\int \frac{d^3k}{((2\pi)^3 2\omega_\mathbf{k})^2}\left[\left|\int e^{-i\mathbf{k}\cdot\mathbf{x}} h(x) d^3x\right|^2 \left(G^{(1)}_{1112}(\bar{\omega},2\bar{\omega},\mathbf{k}) + G^{(2)}_{1211}(\bar{\omega},2\bar{\omega},\mathbf{k})\right)\right. \quad (64)$$
$$\left. + \left|\int e^{i\mathbf{k}\cdot\mathbf{x}} h(x) d^3x\right|^2 \left(G^{(1)}_{2221}(\bar{\omega},2\bar{\omega},\mathbf{k}) + G^{(2)}_{2122}(\bar{\omega},2\bar{\omega},\mathbf{k})\right)\right]$$

where we have introduced a pair of double Laplace transforms

$$G^{(1)}_{ijlm}(\omega_1,\omega_2,\mathbf{k}) = \int_0^\infty \int_0^\infty e^{-\omega_1 t_1} e^{-\omega_2 t_2} g_{ijlm}(t_1,0,t_2,\mathbf{k},\mathbf{k})dt_1 dt_2 \tag{65}$$

and

$$G^{(2)}_{ijlm}(\omega_1,\omega_2,\mathbf{k}) = \int_0^\infty \int_0^\infty e^{-\omega_1 t_1} e^{-\omega_2 t_2} g_{ijlm}(0,t_1,t_2,\mathbf{k},\mathbf{k})dt_1 dt_2. \tag{66}$$

For the right-hand side of Eq. (64) to be positive, it suffices that

$$\mathrm{Re}\left[G^{(1)}_{iiij}(\bar{\omega},2\bar{\omega},\mathbf{k}) + G^{(2)}_{ijii}(\bar{\omega},2\bar{\omega},\mathbf{k})\right] \geq 0,\ i,j=1,2; i\neq j. \tag{67}$$

We have seen that the rotating wave approximation (60) is obeyed exactly by the master Eq. (57). When the transforms $G_{ijlm}$ are derived from the $g_{ijlm}$ implied by Eq. (59), we may compute

$$\begin{aligned}&\mathrm{Re}\left[G^{(1)}_{iiij}(\bar{\omega},2\bar{\omega},\mathbf{k}) + G^{(2)}_{ijii}(\bar{\omega},2\bar{\omega},\mathbf{k})\right]\\ &= \frac{(2\pi)^3 \omega_{\mathbf{k}}\left[1 + \frac{(2\pi)^3 2\omega_{\mathbf{k}}}{\bar{\omega}}(|h_{1i}(\mathbf{k})|^2 + |h_{2i}(\mathbf{k})|^2)\right]}{\omega_{\mathbf{k}}^2 + [\bar{\omega} + (2\pi)^3 2\omega_{\mathbf{k}}(|h_{2i}(\mathbf{k})|^2 - |h_{1i}(\mathbf{k})|^2)]^2}\end{aligned} \tag{68}$$

for $i,j=1,2; i\neq j$. The right-hand side of Eq. (68) is non-negative and we thus conclude that at least for exponential test functions the positive sign of $\mathrm{Re}\, i\Delta_{\mathrm{F,int}}$ is regained for the master equation (57).

**Conclusion**

For the interacting Feynman propagator of scalar electrodynamics, we have shown that the positive sign condition, $\mathrm{Re}\, i\Delta_{\mathrm{F,int}} \geq 0$, may hinge on the reversibility of time evolution. When we switch to reduced dynamics under the weak coupling approximation, this sign is lost for all but extreme cases. This loss arises because we cannot assume that reduced dynamics is reversible without imposing severe restrictions on the Kraus operators that govern time evolution. Fortunately, with another approximation, the rotating wave approximation, we can recover some semblance of a positive sign property by ensuring inequality (67) is true and restricting the test functions to exponentials given by Eq. (55). The field theoretic analog (57) of the quantum optical master equation (56) fulfils this inequality exactly. In contrast to $\mathrm{Re}\, i\Delta_{\mathrm{F,int}}$, the functional $\mathrm{Im}\, i\Delta_F$ is indeterminate.

## Appendix 1

We would like to show that $\mathrm{Im}i\Delta_{F,\mathrm{int}}$ is an indeterminate functional, which we accomplish by looking at the special case $\mathrm{Im}i\Delta_{F,\mathrm{int}} = \mathrm{Im}i\Delta_F$, where $\Delta_F(x,y)$ is the free Feynman propagator. For this purpose, it suffices to demonstrate that

$$\mathrm{Im}i\Delta_F[f] \times \mathrm{Im}i\Delta_F[f'] < 0 \tag{69}$$

for two unequal space-time functions $f$ and $f'$ where

$$\mathrm{Im}i\Delta_F[f] \equiv \mathrm{Im} \int \int f^*(x) i\Delta_F(x,y) f(y) d^4x d^4y. \tag{70}$$

With the use of Eq. (17),

$$\mathrm{Im}i\Delta_F[f] = -\mathrm{Re} \int \int f^*(x) \theta(x_0 - y_0) \int \sin k(x-y) \frac{d^3k}{(2\pi)^3 \omega_\mathbf{k}} f(y) d^4x d^4y \tag{71}$$

where $\omega_\mathbf{k} = \sqrt{\mathbf{k}^2 + m^2}$.

By assuming the spacetime point $x$ to be time-like to enable a Lorentz transformation having only a time coordinate, the integral $\int \omega_\mathbf{k}^{-1} \sin(kx) d^3k$ was computed in [2] (p.159ff) and found to be the sum of two terms. The first singular term vanishes when $x$ is not on the light cone. The second term is also zero outside of the light cone, but inside, a non-zero contribution depending on a first order Bessel function arises due to the massive scalar particle. Such a Lorentz transformation can be avoided by relying on well-known integrals.

The integral with respect to $\mathbf{k}$ on the right-hand side of Eq. (71) may be computed by employing the usual trick of using a coordinate system whose polar axis coincides with $\mathbf{x}$-$\mathbf{y}$. Then, when we adopt spherical coordinates, the polar (inclination) angle and the angle between $\mathbf{k}$ and $\mathbf{x}$-$\mathbf{y}$ are equal, which simplifies the calculation.

$$\mathrm{Im}i\Delta_F[f] = \frac{1}{2\pi^2} \mathrm{Re} \int\int f^*(x) f(y) \theta(x_0 - y_0) \|\mathbf{x} - \mathbf{y}\|^{-1} \times \tag{72}$$
$$\frac{d}{d\|\mathbf{x}-\mathbf{y}\|} \int_0^\infty \frac{\sin\left[(x_0 - y_0)\sqrt{u^2 + m^2}\right]}{\sqrt{u^2 + m^2}} \cos(\|\mathbf{x}-\mathbf{y}\|u) \, du \, d^4x d^4y.$$

According to [23] (integral 3.876.1), with $m, \|\mathbf{x}-\mathbf{y}\|, x_0 - y_0 > 0$,

$$\int_0^\infty \frac{\sin\left[(x_0 - y_0)\sqrt{u^2 + m^2}\right]}{\sqrt{u^2 + m^2}} \cos(\|\mathbf{x}-\mathbf{y}\|u) du$$
$$= \theta(x_0 - y_0 - \|\mathbf{x}-\mathbf{y}\|) \frac{\pi}{2} J_0\left(m\sqrt{(x-y)^2}\right) \tag{73}$$

where $J_0$ is the zeroth order Bessel function and $\theta(u)$ is 1 if $u > 0$ and 0 if $u < 0$. One can also verify that Eq. (73) continues to hold when $x_0 - y_0 = \|\mathbf{x}-\mathbf{y}\|$ provided $\theta(0) = 1/2$. Inserting Eq. (73) into Eq. (72), and using $\frac{dJ_0(z)}{dz} = -J_1(z)$ [24] (9.1.28), and $\frac{d\theta(u)}{du} = \delta(u)$, we arrive at

$$\mathrm{Im}\, i\Delta_F[f] = -\mathrm{Re} \iint f^*(x)f(y)\theta(x_0 - y_0)\left[\frac{\delta((x-y)^2)}{2\pi}\right.$$
$$\left. - \theta((x-y)^2)\frac{m}{4\pi}\frac{J_1(m\sqrt{(x-y)^2})}{\sqrt{(x-y)^2}}\right]d^4x\,d^4y \qquad (74)$$

(cf. [2], p. 160). The quantity in square brackets is, to within a sign, the well-known Jordan-Pauli function [25] (Eq. 2.3.18).

The appearance of the light-like contributing first term inside the square brackets of Eq. (74) arising from a massive scalar particle may be surprising, until we realize that it is exactly canceled with a contribution from the Bessel function term. To this end, we assume a product form of the test functions, $f_\pm(x) = f_{0,\pm}(x_0)g(\mathbf{x})$. Next, we use the three-dimensional convolution theorem in the integration over $\mathbf{x}$ and $\mathbf{y}$. Then, after introducing spherical coordinates, we once again utilize a coordinate system whose polar axis coincides with $\mathbf{x}$ to obtain:

$$\mathrm{Im}\, i\Delta_F[f] = -\frac{2}{(2\pi)^2}\mathrm{Re}\int dx_0 f^*_{0,\pm}(x_0)\int dy_0 f_{0,\pm}(y_0)\theta(x_0 - y_0) \times$$

$$\int \frac{|G(\mathbf{x})|^2}{\|\mathbf{x}\|}\int_0^\infty \sin(\|\mathbf{x}\|r)\left[\frac{\delta((x_0-y_0)^2 - r^2)}{2\pi}\right.$$
$$\left. - \theta((x_0-y_0)^2 - r^2)\frac{m}{4\pi}\frac{J_1(m\sqrt{(x_0-y_0)^2 - r^2})}{\sqrt{(x_0-y_0)^2 - r^2}}\right]r\,dr\,d^3x$$

where $G$ is the Fourier transform of $g$, $G(\mathbf{x}) = \int e^{-i\mathbf{x}\cdot\mathbf{y}}g(\mathbf{y})d^3y$. Using again [24] (9.1.28) yields

$$\mathrm{Im}\, i\Delta_F[f] = -\frac{1}{(2\pi)^3}\mathrm{Re}\int dx_0 f^*_{0,\pm}(x_0)\int dy_0 f_{0,\pm}(y_0)\theta(x_0 - y_0) \times$$
$$\int |G(\mathbf{x})|^2 \int_0^{x_0-y_0} J_0(m\sqrt{(x_0-y_0)^2 - r^2})\cos(\|\mathbf{x}\|r)\,dr\,d^3x, \qquad (75)$$

the boundary term from an integration by parts having exactly canceled the singular term.

At this point, it is convenient to pick

$$f_{0,\pm}(x_0) = \frac{1}{2}\left[\delta(x_0) + i\frac{1}{\pi}\mathcal{P}\frac{e^{\mp i\beta m x_0}}{x_0}\right] \qquad (76)$$

with $\mathcal{P}$ being the Cauchy principal value, $\beta$ any value satisfying $0 < \beta < 1$ and $m > 0$. The peculiar choice of $f_{0,\pm}(x_0)$ is motivated in part by the well-known result ( [26], p. 478) that the Fourier transform of the right-hand side of Eq. (76) is proportional to the simple expression $\theta(y_0 \pm \beta m)$. According to [23] (integral 6.677.6), with $m > 0$,

$$\int_0^{x_0-y_0} J_0\big(m\sqrt{(x_0-y_0)^2 - r^2}\big) \cos(\|\mathbf{x}\|r)\, dr = \frac{\sin\big[(x_0-y_0)\sqrt{\|\mathbf{x}\|^2+m^2}\big]}{\sqrt{\|\mathbf{x}\|^2+m^2}} \tag{77}$$

Inserting (77) into (75) and using the convolution theorem along with Eq. (76), we finally obtain

$$\mathrm{Im}\, i\Delta_F[f_\pm] = \pm \frac{1}{2(2\pi)^4} \int \frac{|G(\mathbf{x})|^2}{\sqrt{\|\mathbf{x}\|^2+m^2}} \ln\left(\frac{\sqrt{\|\mathbf{x}\|^2+m^2} - \beta m}{\sqrt{\|\mathbf{x}\|^2+m^2} + \beta m}\right) d^3x \tag{78}$$

where $m > 0$ and $0 < \beta < 1$. Because the integrand is negative, $\mathrm{Im}\,i\Delta_F[f_-] \times \mathrm{Im}\,i\Delta_F[f_+] < 0$. This shows that $\mathrm{Im}\,i\Delta_F$ is an indeterminate functional.

**Appendix 2**

In this appendix we explore the constraints that Kraus operators must satisfy if the associated dynamical map is invertible. Reduced dynamics is often associated with irreversibility. Indeed, the term irreversible dynamics is sometimes used synonymously with reduced dynamics, which is typically associated with a governing master equation. Dynamical maps can be inverted---a previous approach to reduce noise in quantum computation relied on deconvolutions [27], which are inverses---but this comes at the cost of imposing various restrictions. If we insist that the domain of the inverse is the set of all physical states, $\overline{\mathcal{B}}^+(\mathcal{H})$, as opposed to a circumscribed subset, these restrictions may be too severe to be physically relevant. This means that a dynamical map derived from Eq. (22) and that is onto $\overline{\mathcal{B}}^+(\mathcal{H})$ cannot be inverted for all practical purposes that do not involve trivial or contrived cases.

The relevance to our work is that we have seen that the positive sign property of the Feynman propagator hinges on being able to invert the time evolution propagator under the weak coupling approximation. When the underlying dynamical map is not invertible, the sign can be lost. Thus, it is important to understand why reversibility breaks down.

Along these lines, the authors of [13] used, *inter alia*, contractivity together with Wigner's theorem to prove that a "UDM [universal dynamical map], $\mathcal{E}_{(t_1,t_2)}$, can be inverted by another UDM if and only if it is unitary $\mathcal{E}_{(t_1,t_2)} = \mathcal{U}_{(t_1,t_2)}$." Similarly, it is proven in [19] (Theorem 6.39) that if a unital completely positive map is invertible, then it is a unitary conjugation.

We will see that this result, in which invertibility is used to prove unitarity, also arises as a corollary to the main theorem below, which is obtained using a different approach that relies on the Cauchy-Schwarz inequality for vectors in the Hilbert space $\mathcal{H}$. A similar use of this inequality appears in [28].

We start with two lemmas, the first of which is proved in [13] and is included here for completeness. In this appendix, we suppress the time argument in $\Lambda_t$ and denote by $\mathcal{H}$ a separable Hilbert space.

*Lemma 1*

A convex linear, one-to-one map $\Lambda: \overline{\mathcal{B}}^+(\mathcal{H}) \to \overline{\mathcal{B}}^+(\mathcal{H})$ cannot map a statistical mixture to a pure state.

*Proof:*

A statistical mixture can be written as $p\rho_1 + (1-p)\rho_2$ where $\rho_1, \rho_2 \in \overline{\mathcal{B}}^+(\mathcal{H})$, $\rho_1 \neq \rho_2$ and $0 < p < 1$. Now, suppose it were possible to find a one-to-one map $\Lambda$ such that $\Lambda(p\rho_1 + (1-p)\rho_2) = |\psi\rangle\langle\psi|$, the latter being normalized. Then it is necessary that $\Lambda(\rho_1) = \alpha|\psi\rangle\langle\psi|$ and $\Lambda(\rho_2) = \beta|\psi\rangle\langle\psi|$ such that $p\alpha + (1-p)\beta = 1$. By hypothesis, the trace $\text{Tr}\Lambda(\rho_{1,2}) = 1$ and so $\alpha = \beta = 1$. Hence, $\Lambda(\rho_1) = \Lambda(\rho_2)$ even though $\rho_1 \neq \rho_2$, which is a contradiction for a one-to-one map.

*QED*

The next lemma involves a one-to-one dynamical map whose associated Kraus operators map a particular state to states that are multiples of each other. Since the Kraus linear operators $V_j$ are assumed bounded and everywhere defined in $\mathcal{H}$, the adjoints $V_j^\dagger$ are also bounded and everywhere defined (see Ch. III, Theorem 2.7 of [14]).

*Lemma 2*

Suppose $\Lambda: \overline{\mathcal{B}}^+(\mathcal{H}) \to \overline{\mathcal{B}}^+(\mathcal{H})$ is a one-to-one map given by $\Lambda(\rho) = \sum_k V_k \rho V_k^\dagger$ where the $V_k: \mathcal{H} \to \mathcal{H}$ are bounded linear operators having the property that for any $V_j$ and any $|\phi\rangle \in \mathcal{H}$ satisfying $V_j|\phi\rangle \neq 0$, the relation $V_i|\phi\rangle = \gamma_{ij}(\phi) V_j|\phi\rangle$ $\forall i$ (no Einstein convention) holds, with the $\gamma_{ij}(\phi)$ being complex numbers. Then, there exists at least one $V_{j_*} \in \{V_1, V_2, \dots\}$ such that $V_{j_*}|\psi\rangle \neq 0$ for all normalized states $|\psi\rangle \in \mathcal{H}$.

*Proof:*

Under the hypotheses of the lemma, we will show that a contradiction arises if we assume that for every $V_k$ there exists at least one normalized state $|\psi_{\text{ker}}(k)\rangle$ such that $V_k|\psi_{\text{ker}}(k)\rangle = 0$.

Since $\Lambda$ preserves norm, at least one of the $V_k$ must be non-zero, which we denote by $V_{j_*}$, and a normalized state $|\psi_{\text{nor}}\rangle$ with $V_{j_*}|\psi_{\text{nor}}\rangle \neq 0$ exists. Also, we are presuming that the kernel of $V_k$ includes not only the zero vector but some other vector, for all $k$. In particular, there must be at least one normalized vector $|\psi_{\text{ker}}(j_*)\rangle$ that satisfies $V_{j_*}|\psi_{\text{ker}}(j_*)\rangle = 0$, and which therefore is distinct from $|\psi_{\text{nor}}\rangle$. Due to the linearity of the $V_k$, this distinction must arise from more than a difference in global phase factors.

Now, form the normalized vector

$$|\psi_{\text{sum}}\rangle = \frac{|\psi_{\text{ker}}(j_*)\rangle + |\psi_{\text{nor}}\rangle}{\||\psi_{\text{ker}}(j_*) + \psi_{\text{nor}}\|}. \tag{79}$$

Since $V_{j_*}|\psi_{\text{sum}}\rangle \neq 0$, we have, by hypothesis, $V_i|\psi_{\text{sum}}\rangle = \gamma_{ij_*}(\psi_{\text{sum}})V_{j_*}|\psi_{\text{sum}}\rangle$, $\forall i$. Therefore, we obtain

$$\Lambda(|\psi_{\text{sum}}\rangle\langle\psi_{\text{sum}}|) = \sum_{k=1} V_k |\psi_{\text{sum}}\rangle\langle\psi_{\text{sum}}|V_k^\dagger$$

$$= \sum_{k=1} |\gamma_{kj_*}(\psi_{\text{sum}})|^2 V_{j_*}|\psi_{\text{sum}}\rangle\langle\psi_{\text{sum}}|V_{j_*}^\dagger \tag{80}$$

$$= \frac{V_{j_*}|\psi_{\text{nor}}\rangle\langle\psi_{\text{nor}}|V_{j_*}^\dagger}{\langle\psi_{\text{nor}}|V_{j_*}^\dagger V_{j_*}|\psi_{\text{nor}}\rangle} \tag{81}$$

Likewise, since $V_{j_*}|\psi_{\text{nor}}\rangle \neq 0$, we have $V_i|\psi_{\text{nor}}\rangle = \gamma_{ij_*}(\psi_{\text{nor}})V_{j_*}|\psi_{\text{nor}}\rangle$, $\forall i$. Hence,

$$\Lambda(|\psi_{\text{nor}}\rangle\langle\psi_{\text{nor}}|) = \sum_{k=1} |\gamma_{kj_*}(\psi_{\text{nor}})|^2 V_{j_*}|\psi_{\text{nor}}\rangle\langle\psi_{\text{nor}}|V_{j_*}^\dagger \tag{82}$$

$$= \frac{V_{j_*}|\psi_{\text{nor}}\rangle\langle\psi_{\text{nor}}|V_{j_*}^\dagger}{\langle\psi_{\text{nor}}|V_{j_*}^\dagger V_{j_*}|\psi_{\text{nor}}\rangle} \tag{83}$$

and we see that $\Lambda(|\psi_{\text{sum}}\rangle\langle\psi_{\text{sum}}|) = \Lambda(|\psi_{\text{nor}}\rangle\langle\psi_{\text{nor}}|)$, which is not possible since $|\psi_{\text{sum}}\rangle\langle\psi_{\text{sum}}| \neq |\psi_{\text{nor}}\rangle\langle\psi_{\text{nor}}|$ and we are assuming $\Lambda$ is one-to-one.

We conclude that the assumption that for every $V_k$ there exists at least one normalized state $|\psi_{\text{ker}}(k)\rangle$ such that $V_k|\psi_{\text{ker}}(k)\rangle = 0$ must be false. Consequently, there exists at least one $V_{j_*}$ such that $V_{j_*}|\psi\rangle \neq 0$ for all normalized states $|\psi\rangle$.

*QED*

The main theorem follows and essentially states that the Kraus operators $\{V_1, V_2, \ldots\}$ associated with an invertible dynamical map that is onto all physical states are merely proportional to one another.

*Theorem*

Let $V_j: \mathcal{H} \to \mathcal{H}$ be bounded linear operators and $\Lambda_B: \mathcal{B}_1(\mathcal{H}) \to \mathcal{B}_1(\mathcal{H})$ be a linear map whose restriction $\Lambda(\cdot) = \sum_j V_j \cdot V_j^\dagger$ to the domain $\overline{\mathcal{B}}^+(\mathcal{H}) \subset \mathcal{B}_1(\mathcal{H})$ is onto $\overline{\mathcal{B}}^+(\mathcal{H})$. If $\Lambda$ has an inverse on $\overline{\mathcal{B}}^+(\mathcal{H})$, then $\Lambda(\rho) = N_{j_*}^{-1} V_{j_*} \rho V_{j_*}^\dagger$ where $V_{j_*} \in \{V_1, V_2, \ldots\}$, $N_{j_*} \in (0,1]$ is a normalization factor independent of $\rho \in \overline{\mathcal{B}}^+(\mathcal{H})$, and $N_{j_*}^{-1} V_{j_*}^\dagger V_{j_*} = \mathbb{I}$. The inverse is $\Lambda^{-1}(\cdot) = N_{j_*}^{-1} V_{j_*}^\dagger \cdot V_{j_*}$.

*Proof*:

Convex linearity of $\Lambda$ is inherited from the linearity of $\Lambda_B$, and by hypothesis, the inverse map with domain $\overline{\mathcal{B}}^+$ and range $\overline{\mathcal{B}}^+$ exists, which implies $\Lambda$ and $\Lambda^{-1}$ are one-to-one. Hence, by Lemma 1, $\Lambda$ can neither map a statistical mixture to a pure state nor a pure state to a statistical mixture.

Therefore, for any physical state $\psi$, the image $\Lambda(|\psi\rangle\langle\psi|)$ must be pure, which implies $1 = \text{Tr}\{[\Lambda(|\psi\rangle\langle\psi|)]^2\} = \sum_{k,l}|\langle\psi_k|\psi_l\rangle|^2$ where we have defined $|\psi_k\rangle = V_k|\psi\rangle$. Also, $1 = \text{Tr}\Lambda(|\psi\rangle\langle\psi|) = \sum_k\langle\psi_k|\psi_k\rangle$. Hence, $\sum_{i,j}\left(\langle\psi_i|\psi_i\rangle\langle\psi_j|\psi_j\rangle - |\langle\psi_i|\psi_j\rangle|^2\right) = 0$. But the Cauchy-Schwarz inequality gives $\langle\psi_i|\psi_i\rangle\langle\psi_j|\psi_j\rangle - |\langle\psi_i|\psi_j\rangle|^2 \geq 0$. Thus, $\langle\psi_i|\psi_i\rangle\langle\psi_j|\psi_j\rangle = |\langle\psi_i|\psi_j\rangle|^2$. Equality holds if, and only if, a) $|\psi_j\rangle = 0$ or b) $|\psi_j\rangle \neq 0$ and $|\psi_i\rangle = \gamma_{ij}(\psi)|\psi_j\rangle$, where generally $\gamma_{ij}(\psi)$ are complex numbers that depend on $i,j$ and $\psi$. Consequently, if $V_j \in \{V_1, V_2, \ldots\}$ and $V_j|\psi\rangle \neq 0$, which excludes possibility a), then $V_i|\psi\rangle = \gamma_{ij}(\psi)V_j|\psi\rangle$ for all $i$. The hypotheses of Lemma 2 are met and thus there exists at least one $V_{j_*}$ such that $V_{j_*}|\psi\rangle \neq 0$ for all normalized states $|\psi\rangle \in \mathcal{H}$.

We next show that the $\gamma_{ij_*}(\psi)$ are independent of $\psi$. For all normalized, linearly independent $\phi'$ and $\phi''$, we begin by writing $\|\phi' + \phi''\|V_i\left|\frac{\phi'+\phi''}{\|\phi'+\phi''\|}\right\rangle = V_i|\phi'\rangle + V_i|\phi''\rangle$ and then for the respective vectors use $V_i|\psi\rangle = \gamma_{ij_*}(\psi)V_{j_*}|\psi\rangle$ on both sides to get

$$\left[\gamma_{ij_*}\left(\frac{\phi'+\phi''}{\|\phi'+\phi''\|}\right) - \gamma_{ij_*}(\phi')\right]V_{j_*}|\phi'\rangle + \left[\gamma_{ij_*}\left(\frac{\phi'+\phi''}{\|\phi'+\phi''\|}\right) - \gamma_{ij_*}(\phi'')\right]V_{j_*}|\phi''\rangle = 0. \quad (84)$$

By hypothesis, $\Lambda$ is invertible and therefore one-to-one, and consequently $\{V_{j_*}|\phi'\rangle, V_{j_*}|\phi''\rangle\}$ is linearly independent, for otherwise $V_{j_*}|\phi'\rangle = cV_{j_*}|\phi''\rangle$ where $c$ is a non-zero constant $\Rightarrow$ $V_{j_*}\left(\frac{|\phi'\rangle-c|\phi''\rangle}{\|\phi'-c\phi''\|}\right) = 0$, which would contradict Lemma 2. As a result, we must have

$$\gamma_{ij_*}\left(\frac{\phi'+\phi''}{\|\phi'+\phi''\|}\right) - \gamma_{ij_*}(\phi') = \gamma_{ij_*}\left(\frac{\phi'+\phi''}{\|\phi'+\phi''\|}\right) - \gamma_{ij_*}(\phi'') = 0, \quad (85)$$

which implies $\gamma_{ij_*}(\phi') = \gamma_{ij_*}(\phi'')$. This last equation is also true for any pair of dependent, normalized vectors because if $|\phi''\rangle = c|\phi'\rangle$, then $\left[\gamma_{ij_*}(\phi') - \gamma_{ij_*}(c\phi')\right]V_{j_*}|\phi'\rangle = 0$. This implies $\gamma_{ij_*}(\phi') = \gamma_{ij_*}(\phi'')$ since $V_{j_*}|\phi'\rangle$ is not zero by Lemma 2. We conclude that the $\gamma_{ij_*}(\psi)$ are independent of $\psi$ and we may subsequently omit the argument when writing

$$V_i \cdot V_i^\dagger = |\gamma_{ij_*}|^2 V_{j_*} \cdot V_{j_*}^\dagger. \quad (86)$$

Next, with the help of Eq. (86), we find that for normalized $\psi$,

$$\Lambda(|\psi\rangle\langle\psi|) = N_{j_*}^{-1} V_{j_*}|\psi\rangle\langle\psi|V_{j_*}^\dagger \quad (87)$$

where the normalization constant satisfies

$$N_{j_*}^{-1} = \sum_i |\gamma_{ij_*}|^2$$

$$= \langle\psi|V_{j_*}^\dagger V_{j_*}|\psi\rangle^{-1}. \quad (88)$$

Since the $\gamma_{ij_*}$ are independent of $\psi$, so too is $N_{j_*}$. Noting that $\gamma_{j_*j_*} = 1$, we also have $0 < N_{j_*} \leq 1$. By definition ( [14], p. 395), a density operator $\rho \in \bar{\mathcal{B}}^+(\mathcal{H})$ on a separable Hilbert space $\mathcal{H}$ can be written as $\rho = \sum_{n=1}^{\infty} p_n |n\rangle\langle n|$ where $\{|n\rangle\}$ is an orthonormal basis in $\mathcal{H}$ and $\sum_{n=1}^{\infty} p_n = 1$ with $0 \leq p_n \leq 1$. Along with our assumption that the $V_i$ are bounded and everywhere defined, this enables us to generalize Eq. (87) to $\Lambda(\rho) = N_{j_*}^{-1} V_{j_*} \rho V_{j_*}^\dagger$.

Assuming that $\|\phi\|\|\psi\| \neq 0$, we may now use the polar decomposition and the linearity of $\Lambda_\mathcal{B}$ to obtain

$$\Lambda_\mathcal{B}(|\psi\rangle\langle\phi|) = \begin{cases} \frac{1}{2}\|\psi+\phi\|^2 \Lambda_\mathcal{B}\left(\frac{|\psi+\phi\rangle\langle\psi+\phi|}{\|\psi+\phi\|^2}\right) + \frac{1}{2i}\|\psi-i\phi\|^2 \Lambda_\mathcal{B}\left(\frac{|\psi-i\phi\rangle\langle\psi-i\phi|}{\|\psi-i\phi\|^2}\right) - \frac{1-i}{2}\left[\|\psi\|^2 \Lambda_\mathcal{B}\left(\frac{|\psi\rangle\langle\psi|}{\|\psi\|^2}\right) + \|\phi\|^2 \Lambda_\mathcal{B}\left(\frac{|\phi\rangle\langle\phi|}{\|\phi\|^2}\right)\right] \\ \quad \text{if } \|\psi+\phi\|\|\psi-i\phi\| \neq 0, \\ \frac{1}{2i}\|\psi-i\phi\|^2 \Lambda_\mathcal{B}\left(\frac{|\psi-i\phi\rangle\langle\psi-i\phi|}{\|\psi-i\phi\|^2}\right) - \frac{1-i}{2}\left[\|\psi\|^2 \Lambda_\mathcal{B}\left(\frac{|\psi\rangle\langle\psi|}{\|\psi\|^2}\right) + \|\phi\|^2 \Lambda_\mathcal{B}\left(\frac{|\phi\rangle\langle\phi|}{\|\phi\|^2}\right)\right] \\ \quad \text{if } \|\psi+\phi\| = 0, \\ \frac{1}{2}\|\psi+\phi\|^2 \Lambda_\mathcal{B}\left(\frac{|\psi+\phi\rangle\langle\psi+\phi|}{\|\psi+\phi\|^2}\right) - \frac{1-i}{2}\left[\|\psi\|^2 \Lambda_\mathcal{B}\left(\frac{|\psi\rangle\langle\psi|}{\|\psi\|^2}\right) + \|\phi\|^2 \Lambda_\mathcal{B}\left(\frac{|\phi\rangle\langle\phi|}{\|\phi\|^2}\right)\right] \\ \quad \text{if } \|\psi-i\phi\| = 0. \end{cases} \quad (89)$$

We have taken the trouble to make sure the arguments of $\Lambda_\mathcal{B}$ are in the domain of $\Lambda$ so that we can replace $\Lambda_\mathcal{B}$ by $\Lambda$ and use Eq. (87) to obtain

$$\Lambda_\mathcal{B}(|\psi\rangle\langle\phi|) = N_{j_*}^{-1} V_{j_*} |\psi\rangle\langle\phi| V_{j_*}^\dagger. \quad (90)$$

Using Eq. (89) and the fact that $\Lambda$ preserves norm, we may further obtain

$$\mathrm{Tr}\Lambda_\mathcal{B}(|\psi\rangle\langle\phi|) = \langle\phi|\psi\rangle. \quad (91)$$

We have assumed that $\|\phi\|\|\psi\| \neq 0$ when deriving Eqs. (90) and (91), but the validity of the following equation, which follows therefrom, is apparent for this special case, too, and we can therefore state that

$$N_{j_*}^{-1} \langle\phi| V_{j_*}^\dagger V_{j_*} |\psi\rangle = \langle\phi|\psi\rangle \quad (92)$$

for all $\phi$ and $\psi$, which implies

$$N_{j_*}^{-1} V_{j_*}^\dagger V_{j_*} = \mathbb{I}. \quad (93)$$

By hypothesis, $\Lambda$ has an inverse, $\Lambda^{-1}$, satisfying $\Lambda\Lambda^{-1} = \Lambda^{-1}\Lambda = \mathbb{I}_{\bar{\mathcal{B}}^+}$. From Eqs. (90) and (93), it follows that $N_{j_*}^{-1} V_{j_*}^\dagger \Lambda(\cdot) V_{j_*} = \mathbb{I}_{\bar{\mathcal{B}}^+}(\cdot) \Rightarrow N_{j_*}^{-1} V_{j_*}^\dagger \Lambda(\Lambda^{-1}(\cdot)) V_{j_*} = \Lambda^{-1}(\cdot)$. Hence, $\Lambda^{-1}(\cdot) = N_{j_*}^{-1} V_{j_*}^\dagger \cdot V_{j_*}$.

*QED*

As a corollary to the preceding theorem, we can recover the result of [13] and [19] relating invertibility to unitarity:

Suppose $V_j: \mathcal{H} \to \mathcal{H}$ are bounded linear operators and the convex linear map $\Lambda(\cdot) = \sum_j V_j \cdot V_j^\dagger$ has domain and range $\bar{\mathcal{B}}^+(\mathcal{H})$. If $\Lambda$ is invertible, then $\Lambda(\cdot) = U \cdot U^\dagger$ where $U$ is unitary.

*Proof:*

The preceding theorem tells us that $\Lambda(|\psi\rangle\langle\psi|) = U|\psi\rangle\langle\psi|U^\dagger$ where $U = N_{j_*}^{-1/2} V_{j_*}: \mathcal{H} \to \mathcal{H}$, and Eq. (92) yields $\|U\psi\| = \|\psi\|$ for all $\psi \in \mathcal{H}$. That is, $U$ is isometric. It is also linear because, by hypothesis, so too is $V_{j_*}$. A linear, isometric operator with domain and range $\mathcal{H}$ is unitary (see Ch. III, Theorem 4.3 of [14]), so let us show that $U$ is onto.

We disjointly and exhaustively partition the normalized vectors of $\mathcal{H}$ into equivalence classes such that each class $\mathcal{C}$ consists of normalized vectors equal to each other to within a respective global phase factor. We arbitrarily fix a set of representative vectors $\{\psi\}$ (all normalized) to label the classes with subscripts, $\mathcal{C}_\psi$, such that if $\psi' \in \mathcal{C}_\psi$, then $\psi = e^{-i\theta(\psi')}\psi'$ with $\theta$ real. Obviously, $\psi \in \mathcal{C}_\psi$. There is an isomorphism between these equivalence classes and some physical density operators according to $\mathcal{C}_\psi \mapsto |\psi\rangle\langle\psi|$.

Since $\Lambda(\cdot) = U \cdot U^\dagger$ is onto $\bar{\mathcal{B}}^+$ by hypothesis, for every $|\psi\rangle\langle\psi|$, there exists a physical state, which by Lemma 1 must be a pure state $|\phi\rangle\langle\phi|$, such that $U|\phi\rangle\langle\phi|U^\dagger = |\psi\rangle\langle\psi|$ where $\phi$ and $\psi$ are representative vectors. In terms of equivalence classes, for every $\mathcal{C}_\psi$ there exists a $\mathcal{C}_\phi$ such that $U\mathcal{C}_\phi = \mathcal{C}_\psi$. Thus, the representative vector $\psi$ has a preimage $U|\phi'\rangle = |\psi\rangle$ where $\phi' \in \mathcal{C}_\phi$. Any other $\psi' \in \mathcal{C}_\psi$ is the image of $e^{i\theta}\phi'$ where $\psi' = e^{i\theta}\psi$. Since these considerations apply to all classes, we have shown that all normalized vectors are in the range of $U$. If $\chi$ is an unnormalized, nonzero vector, then its preimage is $\|\chi\|\phi$ where $U|\phi\rangle = |\chi\rangle/\|\chi\|$. Finally, due to linearity of $V_{j_*}$, and therefore of $U$, the zero vector is its own preimage. Hence, $U$ is onto $\mathcal{H}$ and we conclude that $U$ is unitary.

*QED*

This last result is consistent with Lemma 4 of [15] that states that a dynamical map with just one Kraus operator belongs to an extreme ray, and conversely.